\documentstyle[aps]{revtex}
\begin{document}
\title{Bonn Potential and Shell-Model Calculations for $^{206,205,204}$Pb}
\author{L. Coraggio,$^{1,2}$ A. Covello,$^1$ A. Gargano,$^1$ N. Itaco,$^1$
and T. T. S. Kuo$^2$}
\address{$^1$ Dipartimento di Scienze Fisiche, Universit\`a 
di Napoli Federico II, \\ and Istituto Nazionale di Fisica Nucleare, \\
Complesso Universitario di Monte  S. Angelo, Via Cintia - I-80126 Napoli, 
Italy \\
$^2$ Department of Physics, SUNY, Stony Brook, New York 11794}

\date{\today}
\maketitle
\begin{abstract} 
The structure of the nuclei $^{206,205,204}$Pb is studied in terms of
shell model employing a realistic effective interaction derived from the 
Bonn A nucleon-nucleon potential.
The energy spectra, binding energies and electromagnetic properties are 
calculated and compared with experiment. A very good overall agreement is 
obtained. This evidences the reliability of our realistic effective 
interaction and encourages use of modern realistic potentials in shell-model
calculations for heavy-mass nuclei.

\end{abstract}

\draft

The Pb isotopes have long been the subject of great experimental and
theoretical interest. This is of course related to the fact that
$^{208}$Pb is a very good doubly magic nucleus, whose neighbors
are accessible to a variety of spectroscopic studies. This is not
the case for other nuclei in the vicinity of closed shells like the 
$^{100}$Sn and $^{132}$Sn neighbors. These nuclei, in fact, lie well away 
from the valley of stability and only recently our knowledge of their
spectroscopic properties has significantly improved thanks to the
advent of large multidetector $\gamma$-ray arrays. 

From the theoretical point of view the study of nuclei with few valence 
particles or holes provides the best testing ground 
for the basic ingredients of shell-model calculations, especially
as regards the matrix elements of the two-body residual
interaction. In most of the several calculations performed so far
in the lead region, phenomenological potentials have been used 
for the two-body interaction \cite{List,Wang90,Cene97}. As early as some 
twenty-five years ago, however, a realistic effective interaction derived from
the Hamada-Johnston nucleon-nucleon ($NN$) potential\cite{HJ} was employed
in the works of Refs. \cite{Herling72,McGrory75}.
Since that time there has been much progress towards a microscopic approach 
to nuclear structure calculations starting from a free $NN$ potential.
On the one hand, the theoretical framework in which the model-space
effective interaction $V_{\rm eff}$ can be derived from a given $NN$
potential has been largely improved (the main aspects of this derivation
are reviewed in Ref. \cite{Kuo96}). On the other hand, high-quality 
$NN$ potentials have been constructed which give an excellent
description of the $NN$ scattering data. Among these of special
interest for microscopic nuclear structure work are those based
on quantitative meson-theoretic models. A review of the major developments in
this field is given in Ref. \cite{Machl94}.

These improvements have opened the way to a new generation of realistic
shell-model calculations which should assess to which extent 
modern realistic interactions can provide a consistent and accurate
description of nuclear structure phenomena. Until now, however, attention
has been focused on medium-mass nuclei, such as the Sn isotopes and
the $N=82$ isotones \cite{Andr96,Andr97,Cov97,Holt97,Suho98,Holt98}. 
In our own studies \cite{Andr96,Andr97,Cov97}
we considered the $^{100}$Sn neighbors going from $^{102}$Sn to
$^{105}$Sn while for the $N=82$ isotones we were concerned with the
$^{132}$Sn neighbors with two and three valence protons. In both cases
we performed shell-model calculations using a realistic effective
interaction derived from the meson-theoretic Bonn A potential
\cite{Machl87}. The very good agreement between theory and experiment
achieved in these works makes apparent the motivation for the present
study of the $^{206,205,204}$Pb isotopes. These nuclei with two,
three and four holes in the $N$=82-126 shell offer the opportunity
to put to a test our realistic effective interaction in the A=208 region.

In this paper, we assume that $^{208}$Pb is a closed core and let the 
valence neutron holes occupy the six single-hole (s.h.) orbits
$2p_{1/2}$, $1f_{5/2}$, $2p_{3/2}$, $0i_{13/2}$, $1f_{7/2}$, and $0h_{9/2}$.
As regards the energy spacings between the six s.h. levels, we take all
of them from the experimental spectrum of $^{207}$Pb \cite{Martin93}. They are 
(in MeV):
$\epsilon_{f_{5/2}}-\epsilon_{p_{1/2}}=0.570$, $\epsilon_{p_{3/2}}
-\epsilon_{p_{1/2}}=0.898$, $\epsilon_{i_{13/2}}-\epsilon_{p_{1/2}}=1.633$,
$\epsilon_{f_{7/2}}-\epsilon_{p_{1/2}}=2.340$,
and $\epsilon_{h_{9/2}}-\epsilon_{p_{1/2}}=3.414$. 

As in our prior work \cite{Andr96,Andr97,Cov97}, we make use of two-body effective 
interaction derived from the Bonn A free $NN$ potential. The main difference
between the present and earlier calculations is that here we treat neutrons
as valence holes, which implies the derivation of a hole-hole effective
interaction. This was obtained using a $G$-matrix formalism, including 
renormalizations from both core polarization and folded diagrams. We have 
chosen the Pauli exclusion operator $Q_2$ in the $G$-matrix equation,
$$G(\omega)=V+V Q_2 {{1} \over
{\omega-Q_2TQ_2}} Q_2G(\omega), \eqno (1)$$
as specified \cite{Kuo96} by ($n_1, n_2, n_3$) = (22, 36, 66) for the neutron
orbits and ($n_1, n_2, n_3$) = (16, 28, 66) for the proton orbits.  Here $V$
represents the $NN$ potential, $T$ denotes the two-nucleon kinetic energy,
and $\omega$ is the so-called starting energy. We employ a matrix inversion
method to calculate the above $G$ matrix in an essentially 
exact way \cite{Krenc76}. In the calculation of the effective interaction 
we take the so-called $\hat Q$-box \cite{Kuo96} to be composed of $G$-matrix
diagrams through second order in $G$. They are the seven first- and second-order 
diagrams considered in Ref. \cite{Shurp83} with the particle lines
replaced by hole lines. This brings about changes in the phase factors and
off-shell energy variables.
Since in $^{208}$Pb neutrons and protons have different closed shell cores,
$Z=82$ and $N=126$, respectively, in the calculation of $V_{\rm eff}$ we use an 
isospin uncoupled representation, where protons and neutrons are treated 
separately. For the shell-model oscillator $\hbar \omega$ we use the value 
6.88 MeV, as obtained from the expression
$\hbar \omega = 45 A^{-{ 1 \over 3}} - 25 A^{-{2 \over 3}}$ for A=208.

The experimental \cite{Helmer90,Rab93} and theoretical spectra 
of $^{206}$Pb and $^{205}$Pb
are compared in Figs. 1 and 2, where we report all the calculated and
experimental levels up to 2.5 and 1.5 MeV for the former and the
latter, respectively. In the higher-energy region we only compare the
calculated high-spin states with the observed ones. As regards $^{204}$Pb,
all experimental \cite{Schmorak94} and calculated levels up to 2.0 MeV 
are reported in Fig. 3
while high-spin states are shown in Fig. 4. From Figs. 1-3 we see that
a very good agreement with experiment is obtained for the low-energy
spectra. In particular, in each of the three nuclei the theoretical
level density reproduces remarkably well the experimental one. Note
too that each state of a given $J^\pi$ in any of three calculated
spectra has its experimental counterpart, with a few exceptions. In fact,
as may be seen in Fig. 2, the ${5 \over 2}^-$, 
$({{3 \over 2}, {1 \over 2}})^-$, and $({{9 \over 2}, {7 \over 2}})^-$
states observed at 1.265, 1.374 
and 1.499 MeV in $^{205}$Pb cannot be safely identified with levels 
predicted by the theory. As regards $^{204}$Pb, we find the $0_4^+$
state at 1.954 MeV while the experimental one, which is not reported
in Fig. 3, lies at 2.433 MeV. It should be mentioned, however, that
the theory predicts four more $0^+$ states in the energy interval
2.2--2.6 MeV. Aside from these uncertainties, the agreement between
calculated and experimental spectra is such as to allow us to
identify experimental states with no firm or without spin-parity
assignment. For $^{206}$Pb our results suggest that the observed
levels at 2.197 and 2.236 MeV have $J^\pi = 3^+$ and $1^+$,
respectively. As for $^{205}$Pb, we predict $J^\pi$ =
${ {1 \over 2}^-}$  and ${ {3 \over 2}^-}$  for the experimental
levels at 0.803 an 0.998 MeV.

Regarding the quantitative agreement between our results and experiment,
the discrepancy for the $2^+_1$ states in $^{206}$Pb and $^{204}$Pb
is only about 40 keV, while all other excited states in the low-energy
spectra of both nuclei lie about 200 keV below the experimental ones.
The rms deviation $\sigma$ \cite{sigma} is 207 and 216 keV for
$^{206}$Pb and $^{204}$Pb, respectively. The agreement with experiment 
is even better for $^{205}$Pb. In this case the $\sigma$ value is
74 keV, excluding the three above mentioned states, for which we
have not attempted any identification.

Concerning the high-spin states in $^{206}$Pb and $^{205}$Pb, from
Figs. 1 and 2 we see that they are also well described by the
theory. In $^{204}$Pb the agreement between theory and experiment
is rather worse for the states lying above 4.3 MeV excitation energy,
the largest discrepancy being about 400 keV for the ${16}^+_2$ state. 

We have also calculated the ground-state binding energies (relative to
$^{208}$Pb). The mass excess value for $^{207}$Pb needed for absolute
scaling of the s.h. levels was taken from \cite{Audi93}. We find
$E_b$($^{206}$Pb)=-14.240, $E_b$($^{205}$Pb)=-22.147, and 
$E_b$($^{204}$Pb)=-28.927 MeV, to be compared with the experimental
values -14.106(6), -22.194(6), and -28.925(6) MeV \cite{Audi93},
respectively.

Let us now come to the electromagnetic observables. Concerning the
magnetic properties, we have specified the effective $M1$ operator
in the following way. Five s.h. matrix elements have been
determined from the measured magnetic moments and $M1$ transition 
rates in $^{207}$Pb. The available experimental information
regards the moments of the ${{1 \over 2}^-}$, ${{5 \over 2}^-}$,
and ${{3\over 2}^-}$ states \cite{Rag89,Kap70} and the 
$B(M1; {{3 \over 2}^-} \rightarrow {{1 \over 2}^-})$ and
$B(M1; {{3 \over 2}^-} \rightarrow {{5 \over 2}^-})$
\cite{Martin93}. The effective $i_{13/2}$ $M1$ operator has
been determined from the magnetic moment of the $12^+$ state in
$^{206}$Pb which arises from the $(i_{13/2})^{-2}$ configuration.
For the remaining matrix elements, we have used the bare operator 
quenched by the factor 0.6. In this way, the $M1$ operator
was specified by nine s.h. matrix elements.
In Table I we compare the experimental magnetic moments
in $^{206,205,204}$Pb \cite{Rag89} with the values calculated with both
the bare operator and the effective $M1$ operator specified above.
We see that the latter values are in very good agreement with 
experiment, most of them falling within the error bars. 
The only significant discrepancy is the sign of the magnetic moment
of the $6^-$ state in $^{206}$Pb. It should be noted that this
disagreement was also found in Ref. \cite{McGrory75}, where
the difficulty to understand the measured positive value is evidenced.
We fully agree with the conclusion of the above work and think that a new
measurement of this magnetic moment is most desirable.
It is worth mentioning that, as can be easily verified from Table I,
no state-independent quenching of the bare operator can lead to a 
satisfactory agreement.
Only one $B(M1)$ value is known. This is the $B(M1; 6^- \rightarrow 7^-)$ 
in $^{206}$Pb which has been measured to be 0.045(13) W.u. 
\cite{Martin93}. Our calculated value is 0.132 W.u.

As regards the calculation of the $E \lambda$ observables, we have used  
an effective neutron hole charge $e_n^{\rm eff}=0.82 e$. This has been
obtained from the observed $B(E2; {{5 \over 2}^-} \rightarrow {{1 \over 2}^-})$
in $^{207}$Pb \cite{Martin93}.
In Tables II and III we compare the calculated quadrupole moments
and $E \lambda$ transition rates with the experimental ones 
\cite{Rag89,Helmer90,Rab93,Lind76,Schmorak94}. Generally, the agreement is very
good, the main discrepancy regarding the sign of the quadrupole moment
of the $2^+$ state in $^{204}$Pb.

In summary, we have presented here the results of a shell-model study of the 
neutron hole isotopes $^{206,205,204}$Pb, where use has been made of an 
effective two-hole interaction derived from the Bonn A nucleon-nucleon 
potential. We have shown that a large number of experimental data 
regarding the three nuclei considered are very well reproduced
by the theory. It should be emphasized that these are the first shell-model
calculations for heavy-mass nuclei in which the effective interaction 
is derived from a modern $NN$ potential by means of a $G$-matrix folded 
diagram method. In fact, as already mentioned, the earlier realistic
calculation of Ref. \cite{McGrory75} made use of an effective interaction 
derived from the Hamada-Johnston potential and including only the bare 
interaction and the core polarization (or bubble) diagram. In addition, 
to obtain good agreement with experiment, the bubble diagram matrix elements 
were multiplied by the single empirical constant 0.75.
The same effective interaction has been recently used \cite{Raman96} 
to describe the results of a detailed experimental study of $^{206}$Pb via
the $^{205}$Pb($n, \gamma$) reaction. 

We may conclude that our present results, which are quite consistent
with those obtained for nuclei around $^{100}$Sn and $^{132}$Sn,
provide further insight into the role of modern realistic 
interactions in nuclear structure calculations, evidencing,
in particular, the merit of the Bonn potential.

\acknowledgments
This work was supported in part by the Italian Ministero dell'Universit\`a
e della Ricerca Scientifica e Tecnologica (MURST) and by the U.S. DOE Grant
No. DE-FG02-88ER40388. One of us (L.C.) wishes to acknowledge the hospitality
received while at SUNY and also thank the Angelo Della Riccia Foundation 
for financial support.

\begin{figure}
\caption{Experimental and calculated spectrum of $^{206}$Pb.}
\label{fig.1}
\end{figure}
\begin{figure}
\caption{Experimental and calculated spectrum of $^{205}$Pb.}
\label{fig.2}
\end{figure}
\begin{figure}
\caption{Experimental and calculated low-energy spectrum of $^{204}$Pb.}
\label{fig.3}
\end{figure}
\begin{figure}
\caption{Experimental and calculated high-spin states in $^{204}$Pb.}
\label{fig.4}
\end{figure}

\begin{table}
\caption {Calculated and experimental magnetic moments (in n.m.) in
$^{206,205,204}$Pb. The theoretical values have been obtained by
using (a) an effective $M1$ operator (se text for details), and
(b) the free $M1$ operator.}

%\mediumtext
%\setdec 0.00

\begin{tabular}{llccc}
 Nucleus & $J^{\pi}$ & \multicolumn{3}{c} {$\mu$} \\
 ~~~&~~~& Expt. & Calc.(a) & Calc.(b) \\
\tableline
 $^{206}$Pb &       $2_1^+$         & $\leq 0.030$ & 0.057  & 0.340  \\
 ~~~~~~~~~  &       $7_1^-$         & -0.1519 (28) & -0.277 & -0.736 \\
 ~~~~~~~~~  &       $6_1^-$         & 0.78 (42)    & -1.20  & -2.02  \\
 ~~~~~~~~~  &       $12_1^+$        & -1.795 (22)  & -1.794 & -3.532 \\
 $^{205}$Pb & $(\frac{5}{2}^-)_1  $ & 0.7117 (4)   &  0.695 &  1.185 \\
 ~~~~~~~~~~ & $(\frac{13}{2}^+)_1 $ & -0.975 (40)  & -0.897 & -1.794 \\
 ~~~~~~~~~~ & $(\frac{25}{2}^-)_1 $ & -0.845 (14)  & -1.010 & -2.564 \\
 ~~~~~~~~~~ & $(\frac{33}{2}^+)_1 $ & -2.442 (83)  & -2.467 & -4.856 \\
 $^{204}$Pb &       $2_1^+$         & $< 0.02$     &  0.04  &  0.30  \\
 ~~~~~~~~~  &       $4_1^+$         & 0.225 (4)    &  0.306 &  0.856 \\
\end{tabular}
\label{Table I}
\end{table}

\begin{table}
\caption {Calculated and experimental electric quadrupole moments
(eb) in $^{206,205,204}$Pb.}

%\mediumtext
%\begin{table}
%\setdec 0.00

\begin{tabular}{llcc}
Nucleus & $J^{\pi}$ & \multicolumn{2}{c} {Q} \\
 ~~~&~~~& Expt. & Calc. \\
\tableline
 $^{206}$Pb &       $2_1^+$         & 0.05 (9)   &  0.26  \\
 ~~~~~~~~~  &       $7_1^-$         & 0.33 (5)   &  0.37  \\
 ~~~~~~~~~  &       $12_1^+$        & 0.51 (2)   &  0.46  \\
 $^{205}$Pb & $(\frac{5}{2}^-)_1  $ & 0.226 (37) &  0.164 \\
 ~~~~~~~~~~ & $(\frac{13}{2}^+)_1 $ & 0.30 (5)   &  0.35  \\
 ~~~~~~~~~~ & $(\frac{25}{2}^-)_1 $ & 0.63 (3)   &  0.55  \\
 $^{204}$Pb &       $2_1^+$         & 0.23 (9)   & -0.11  \\
 ~~~~~~~~~  &       $4_1^+$         & 0.44 (2)   &  0.32  \\
\end{tabular}
\label{Table II}
\end{table}

\begin{table}
\caption {Calculated and experimental $B(E \lambda)$ (in W.u.) in
$^{206,205,204}$Pb.}

%\mediumtext
%\begin{table}
%\setdec 0.00

\begin{tabular}{lllcc}
 Nucleus & $J^{\pi}_i \rightarrow J^{\pi}_f$ & $\lambda$ & \multicolumn{2}{c}
{B(E$\lambda$)} \\
 ~~~&~~~&~~~& Expt. & Calc. \\
\tableline
 $^{206}$Pb & $2_1^+ \rightarrow 0_1^+$ & 2 & 2.85 (3)   & 2.64  \\
 ~~~~~~~~~  & $6_1^- \rightarrow 7_1^-$ & 2 & $\leq 0.4$ & 0.05  \\
 ~~~~~~~~~  & $7_1^- \rightarrow 4_2^+$ & 3 & 0.28 (4)   & 0.11   \\
 ~~~~~~~~~  & $7_1^- \rightarrow 4_1^+$ & 3 & 0.36 (6)   & 0.21  \\
 $^{205}$Pb & $(\frac{25}{2}^-)_1 \rightarrow (\frac{21}{2}^-)_1$ & 2 & 
 0.62 (2)     & 0.60   \\
 ~~~~~~~~~~ & $(\frac{33}{2}^+)_1 \rightarrow (\frac{29}{2}^+)_1$ & 2 & 
 0.63 (21)    & 0.60   \\
 ~~~~~~~~~~ & $(\frac{13}{2}^+)_1 \rightarrow (\frac{7}{2}^-)_1$  & 3 & 
 0.00198 (22) & 0.0002 \\
 ~~~~~~~~~~ & $(\frac{25}{2}^-)_1 \rightarrow (\frac{19}{2}^+)_1$ & 3 & 
 0.088 (8)    & 0.008  \\
 ~~~~~~~~~~ & $(\frac{33}{2}^+)_1 \rightarrow (\frac{27}{2}^-)_1$ & 3 & 
 0.15 (3)     & 0.01   \\
 ~~~~~~~~~~ & $(\frac{33}{2}^+)_1 \rightarrow (\frac{29}{2}^-)_1$ & 3 & 
 0.17 (2)     & 0.01   \\
 $^{204}$Pb & $2_1^+ \rightarrow 0_1^+$ & 2 & 4.65 (6)     & 3.28    \\
 ~~~~~~~~~  & $4_1^+ \rightarrow 2_1^+$ & 2 & 0.00382 (14) & 0.08 \\
 ~~~~~~~~~  & $0_2^+ \rightarrow 2_1^+$ & 2 & $\leq 0.80$  & 0.01    \\
 ~~~~~~~~~  & $4_1^+ \rightarrow 0_1^+$ & 4 & 2.5 (5)      & 3.3     \\
\end{tabular}
\label{Table III}
\end{table}


\begin{references}
\bibitem{sigma} We define $\sigma= \{ (1/N_d) \sum_i [E_{\rm expt}(i)-
E_{\rm calc}(i)]^2 \}^{1\over2}$, where $N_d$ is the number of data.
\end{references}
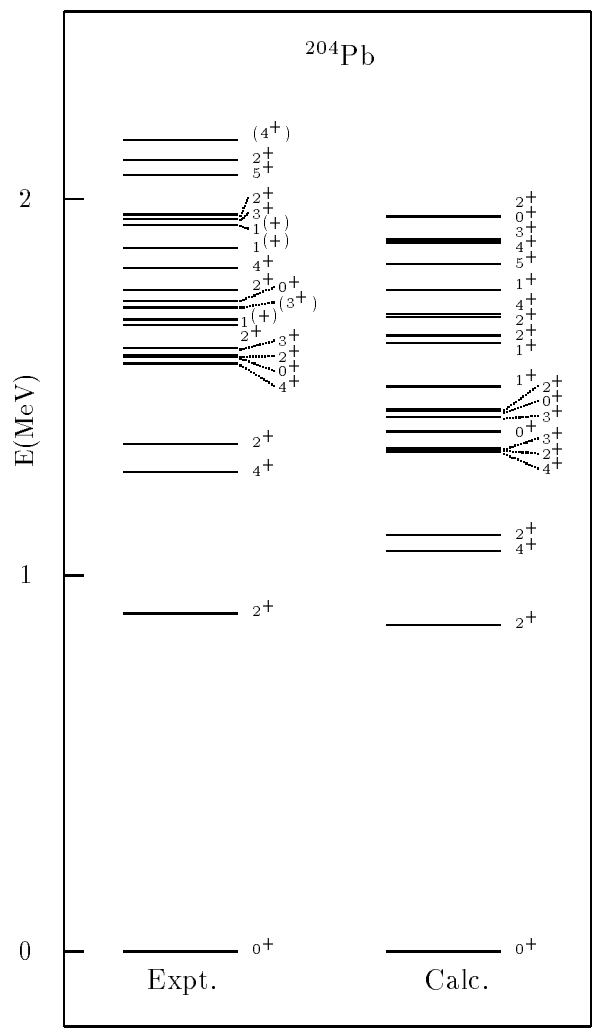
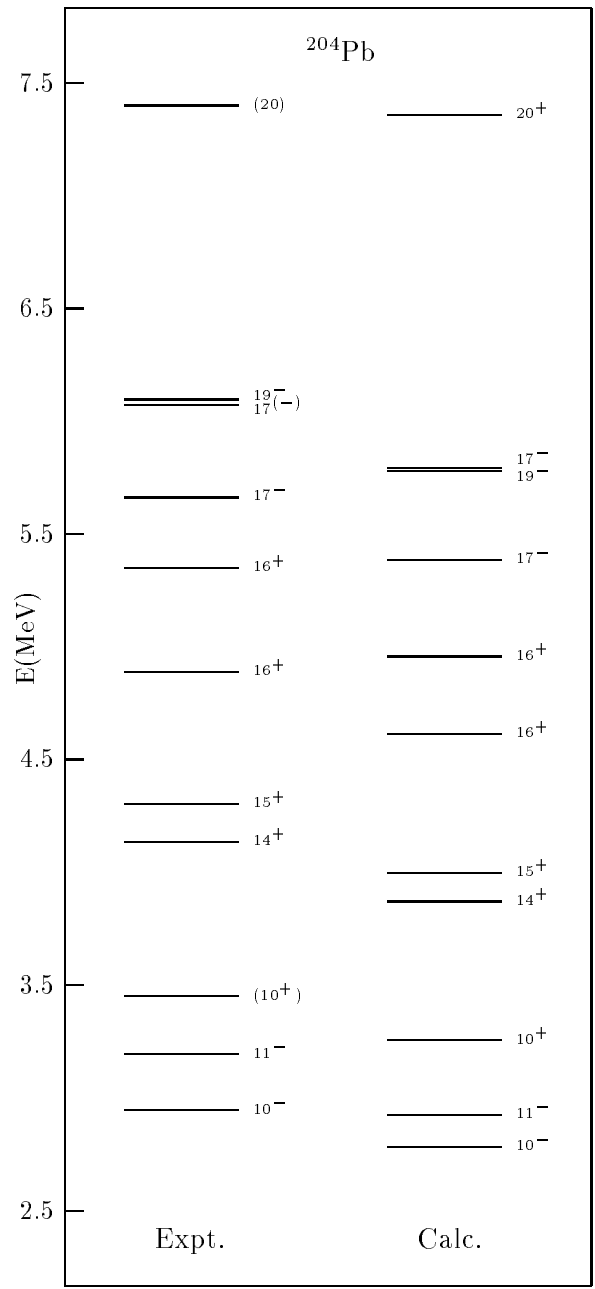
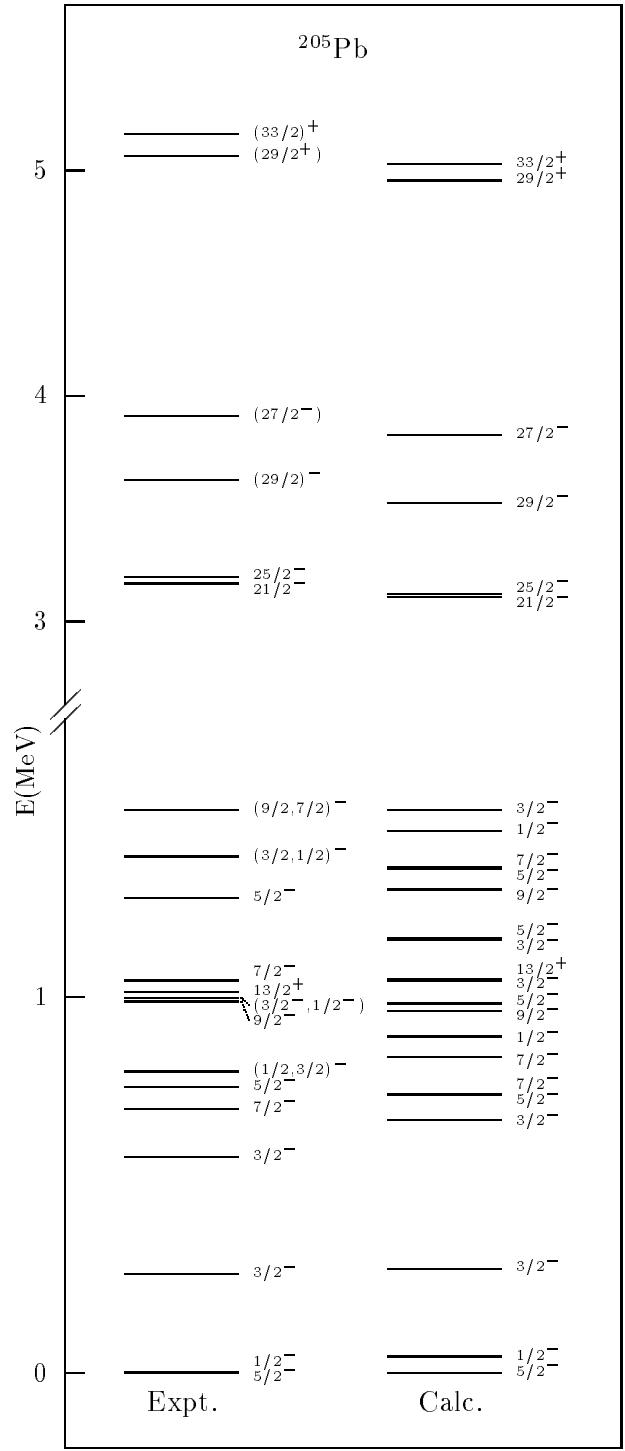
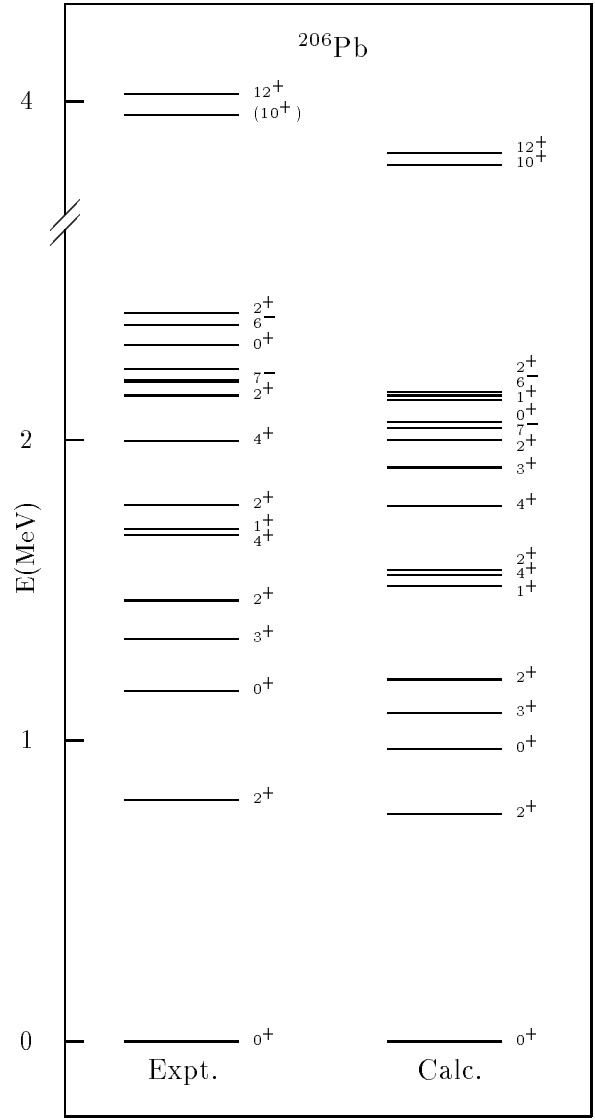
\end{document}